\documentclass[aip,amsmath,amssymb,  reprint]{revtex4-1}
\usepackage{graphicx}  
\usepackage{amsmath}
\usepackage{bm}        
\usepackage{amssymb}   

\usepackage{float}
\usepackage{import}

\usepackage{color}

\begin{document}

\title{Infinite-pressure phase diagram of binary mixtures of (non)additive hard disks}

\author{Etienne Fayen}
\affiliation{Universit\'e Paris-Saclay, CNRS, Laboratoire de Physique des Solides, 91405 Orsay, France}
\author{Anuradha Jagannathan}
\affiliation{Universit\'e Paris-Saclay, CNRS, Laboratoire de Physique des Solides, 91405 Orsay, France}
\author{Giuseppe Foffi}
\email{giuseppe.foffi@u-psud.fr}
\affiliation{Universit\'e Paris-Saclay, CNRS, Laboratoire de Physique des Solides, 91405 Orsay, France}
\author{Frank Smallenburg}
\email{frank.smallenburg@u-psud.fr}
\affiliation{Universit\'e Paris-Saclay, CNRS, Laboratoire de Physique des Solides, 91405 Orsay, France}

\date{\today}
\begin{abstract}
One versatile route to the creation of two-dimensional crystal structures on the nanometer to micrometer scale is the self-assembly of colloidal particles at an interface. Here, we explore the crystal phases that can be expected from the self-assembly of mixtures of spherical particles of two different sizes, which we map to (additive or non-additive) hard-disk mixtures. We map out the infinite-pressure phase diagram for these mixtures, using Floppy Box Monte Carlo simulations to systematically sample candidate crystal structures with up to 12 disks in the unit cell. As a function of the size ratio and number ratio of the two species of particles, we find a rich variety of periodic crystal structures. Additionally, we identify random tiling regions to predict random tiling quasicrystal stability ranges. Increasing non-additivity both gives rise to additional crystal phases and broadens the stability regime for crystal structures involving a large number of large-small contacts, including random tilings. Our results provide useful guidelines for controlling the self-assembly of colloidal particles at interfaces.
\end{abstract}
\maketitle

\section{Introduction}

The self-assembly of colloidal particles into organized crystals provides an elegant and versatile route for the creation of materials with well-controlled structure on the nanometer to micrometer scale. By varying e.g. the shape, size distribution, and surface properties of the building blocks, a stunning variety of crystal structures can be obtained \cite{glotzer2007anisotropy, vanmaekelbergh2011self, sacanna2013engineering, boles2016self}. Even in the seemingly simple case of spherical particles self-assembling at a two-dimensional interface or substrate, mixing particles of different sizes has been demonstrated to lead to a variety of crystalline \cite{kim2005rapid, yu2010co, zhang2010colloidal, Dong2011, zhang2012fabrication, lotito2017approaches} and even quasicrystalline \cite{talapin2009, ye2017quasicrystalline} structures.

When spherical particles self-assemble at an interface, it can be useful to consider an effective problem where they essentially act as two-dimensional disks, forming a two-dimensional ordered structure~\cite{ Dong2011, Ye2015}. Which crystal structure is selected for self-assembly depends on the (effective) interactions between these disks. In the simplest scenario -- a single species of particles interacting via a short-ranged interaction potential -- the inevitable outcome is a hexagonal lattice. However, mixing two types of particles with differing interactions already leads to an impressive complexity. Binary systems with soft repulsive interactions, due to e.g. charge or dipolar forces, stabilize a wide variety of binary crystal structures (see e.g. \cite{assoud2007stable, assoud2008binary, fornleitner2008, fornleitner2009ordering, Law2011}). Even simple hard disks, with no interactions beyond a hard-core exclusion, are predicted to have a rich and varied phase diagram \cite{likos1993, uche2004}, containing periodic crystals as well as lattice gases and random tilings. Moreover, as these phases are all expected to be stable in the limit of high packing fractions, they will be relevant for any high-density system where the interactions include a hard repulsive core.
As such, understanding the phase behavior of mixtures of hard disks provides useful insights into a wide range of (quasi-)two-dimensional self-assembly processes. Finally, the existence and control of  random tiling phases can help to rationalize the existence and behaviour of quasicrystalline  self-assembly~\cite{ye2017quasicrystalline} -- a new emerging field in soft matter~\cite{talapin2009, dotera2014, wang2018}.

In this work, we explore the crystal structures formed by mixtures of hard disks of two different sizes, in 2D, considering both additive and non-additive mixtures. A phase diagram for additive i.e. non-overlapping hard disks, at infinite pressure has been proposed by Likos and Henley \cite{likos1993}, based on a large set of candidate structures built explicitly using clever heuristics and deformation arguments. However, since there are an infinite number of possible crystal structures, it is  impossible in practice to make sure that no phase of even higher compacity has been missed. Here, we systematically detect candidate crystal structures using so-called \emph{Floppy Box Monte Carlo} simulations\cite{filion2009, degraaf2012}. While this method still inevitably leaves room for missed crystal phases, we find both new stable crystals and several better-packed deformations that were not considered in earlier work\cite{likos1993}. This leads to an updated phase diagram, which also includes two new regions where random tiling phases (and the associated quasicrystals) are expected. We then explore how non-additivity, i.e. allowing the possibility that disks may overlap, affects the phase diagram. As explained in the next section, non additivity allows to mimic possible 3D effects that may occur when different types of colloidal particles with different sizes self-assemble at an interface and float at different levels (see Fig.~\ref{fig:NAHD_interpretation}).
We find three new crystal phases which are only stable for finite non-additivity, and we map out how the stability of the other structures shifts as non-additivity increases. Our results show that non-additivity has a drastic effect on the phase diagram, strongly increasing the stability range of crystals (and quasicrystals) that have a large number of contacts between large and small disks.

In the remainder of this paper, we first present the (non-)additive hard disks model (Section \ref{sec:model}). In Section \ref{sec:methods} we outline the numerical simulations used to sample candidate structures and construct the phase diagrams. Sections \ref{sec:HD} and \ref{sec:NAHD} describe the main features of the phase diagrams of binary mixtures of additive and non-additive hard disks respectively. Finally, we discuss our findings and conclude the paper in Section \ref{sec:discussion}. 

\section{(Non)additive hard-disk model}
\label{sec:model}
We consider, in two dimensions, mixtures of large and small hard disks (HD) with diameters $\sigma_L$ and $\sigma_S$ respectively. 
Such mixtures are characterised by the size ratio of the large and small disks $q=\sigma_S/\sigma_L$ and the number fraction of small disks $x_S=N_S/(N_L+N_S)$, with $N_L$ and $N_S$ the number of large and small disks respectively. 

The interaction potentials between two disks $u_{ij}$ ($i, j = L \text{ or } S$ for large and small disks respectively) a distance $r$ apart, is given by 
\begin{equation}
\label{eq:pot}
u_{ij}(r) = \begin{cases}
    \infty &\text{ if } r < \sigma_{ij}\\
    0 &\text{ otherwise.}
    \end{cases}
\end{equation}
Where $\sigma_{SS} =\sigma_S$ and $\sigma_{LL} =\sigma_L$. The interspecies diameter is defined as
\begin{equation}
    \sigma_{LS} = \sigma_{SL} = \left(1-\Delta \right)\frac{\sigma_S+\sigma_L }{2}
\end{equation}
where $\Delta$ is the so-called \emph{non-additivity parameter}. 
When $\Delta = 0$, the model is called \emph{additive} and the disks behave like standard hard disks that cannot overlap. In this paper we will study the case of \textit{negative additivity}, i.e. $\Delta > 0$ that will be implied for the rest of the paper.

\begin{figure}
    \centering
    \includegraphics[width=1.\linewidth]{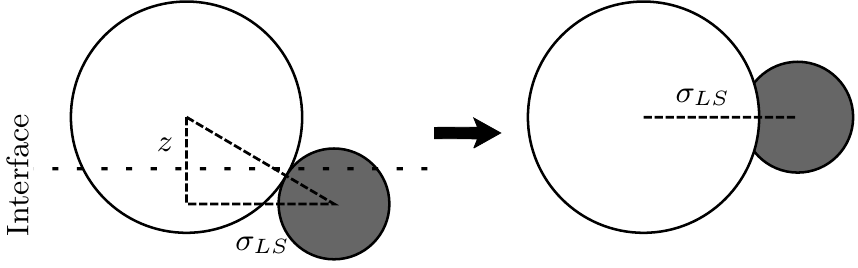}
    \caption{Hard disks can be interpreted as hard spheres floating at an interface. If the two species do not float at the same level (left figure), the corresponding 2D "top view" exhibits overlap between large and small disks (right figure).}
    \label{fig:NAHD_interpretation}
\end{figure}

However, 2D (or quasi-2D) superlattices are often formed out of 3D spherical particles, self-assembling at an interface \cite{kim2005rapid, yu2010co, zhang2010colloidal, Dong2011, zhang2012fabrication, lotito2017approaches}. In principle, large and small nanoparticles may have different surface properties, leading to different wetting angles with the solvent. As a result, the two species may float at different levels with respect to the interface.
In such cases, small particles can partially slide below (or above) the large ones, as illustrated in Fig.~\ref{fig:NAHD_interpretation}~\cite{salgado2015non}.
This 3D effect can be accounted for by allowing large and small disks to overlap slightly.
This is achieved by introducing some degree of negative additivity as discussed above. From simple geometrical considerations, one can show that if the floating levels of the two species are offset by a distance $z$, the non-additivity parameter is given by $\Delta=1-\sqrt{1-{4 z^2}/{(\sigma_S+\sigma_L)^2}}$.
Particles of the same species, since they lie at the same level with respect to the interface, are not allowed to overlap: their minimal approach distances are unchanged. 

We note that other sources of non-additivity can be readily present in real-world self-assembling systems. Particles that interact via softer interactions (due to charge, dipolar interactions, ligand coatings, etc.), may, depending on the overall packing fraction, favor configurations where the favored distance between different pairs of particles behaves non-additively. Non-additive interactions have been shown to significant impact the phase behavior of mixtures of particles \cite{dijkstra1998phase, louis2000crystallization,saija2002monte, widmer-cooper2011, salgado2015non}, and hence are likely to be an important factor in predicting the self-assembly of hard-disk mixtures.

In this work, we focus on finding the stable crystal structures in the limit of high pressures. In this limit, the most stable phase for any given combination of composition $x_S$, size ratio $q$, and non-additivity parameter $\Delta$ is the best-packing phase: the one with the lowest volume per particle $v$. 

\section{Methods}
\label{sec:methods}
To construct a phase diagram, one first needs to know the phases that compete for stability, i.e. candidate packings of disks with a given size ratio and composition. 

In order to systematically generate candidate crystal structures, we use the so-called \emph{Floppy-box Monte Carlo} (FBMC) simulation method \cite{filion2009}. In this method, we simulate a small number of particles (up to 12) in a periodic box at slowly increasing pressures and compress it until a dense packing is reached. To accommodate all unit-cell shapes, Monte Carlo moves are included which deform the simulation box. At the end of the simulation, a quench to infinite pressure is performed to freeze the configuration, which is then taken as a candidate structure. 

\begin{figure*}
    \centering
    \includegraphics[width=0.8\linewidth]{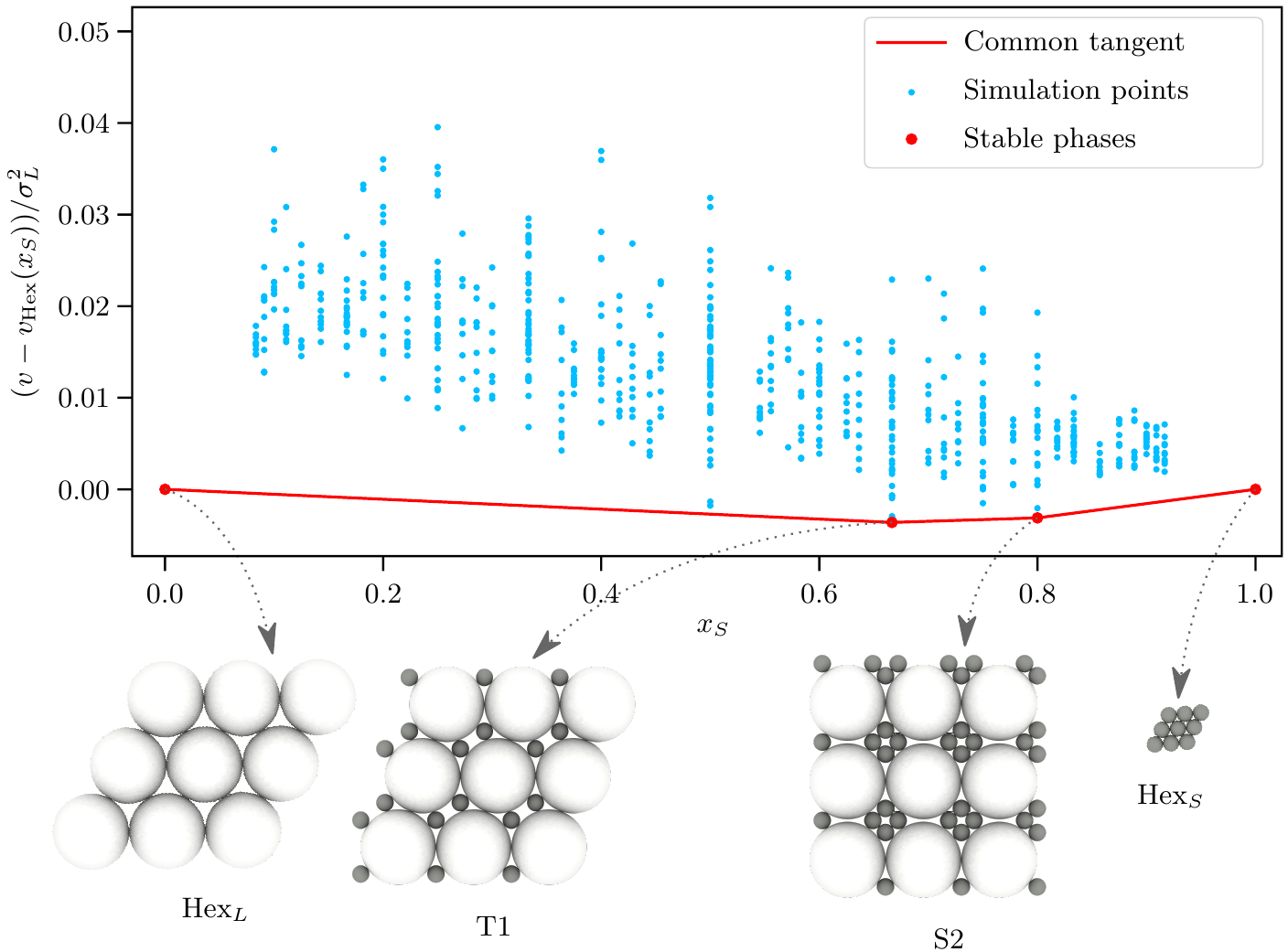}
    \caption{Common tangent construction at size ratio $q=0.23$. On the vertical axis, we subtracted the volume per particle $v_\mathrm{Hex}(x_S)$ of the coexistence of hexagonal phases of large and small disks. Blue dots represent candidate crystal structures found by 50 FBMC simulations for each of the 66 compositions $LiSj$ with $i,j\geq1$ and $i+j\leq 12$. Red dots correspond to stable pure crystals and arrows point to snapshots of the FMBC output (unit cells are repeated 9 times). The red line indicates, for all intermediate fraction of small disks, the volume per particle of the stable coexistence.}
    \label{fig:common_tangent}
\end{figure*}

In our simulations, we considered all possible compositions with up to 12 disks in the simulation box and at least one large and one small disk. For larger numbers of particles, the method becomes less reliable, as the number of possible arrangements increases very rapidly with the number of particles. Nonetheless, this has proven to be an effective method for systematically finding unit cells of complex crystal structures in a large variety of systems \cite{torquato2009dense, degraaf2012, marechal2012freezing, bianchi2012predicting, vissers2013predicting, staneva2015role, gabrielse2017low, shen2019symmetries}, as long as the unit cell is not too large. Since the number of particles in the box is small, simulations are fast, allowing us to produce at least 50 candidate structures for each composition and size ratio we investigated. Note that typically, the most efficiently packed crystal structures are found multiple times in these 50 independent runs.

Equipped with a set of candidate structures, we used an approach that is  analogous to a \emph{common tangent construction} to determine the relative stability of the corresponding phases. In particular, for a given size ratio $q$, we plot each obtained candidate structure as a point in the $(x_S - v)$-plane, where $v$ is the volume per particle. 
At any $x_S$, the stable state is the crystal phase, or coexistence of two crystal phases, which has the lowest volume per particle. For a coexistence of two phases $\alpha$ and $\beta$, involving a fraction $n_\alpha$ and $n_\beta=1-n_\alpha$ of all particles respectively, the overall fraction of small disks is $x_S=n_\alpha x_{S,\alpha} + n_\beta x_{S,\beta}$.
The volume per particle $v$ of the coexistence is also linear in the number fraction of particles involved each of the phases. 
Therefore, in the $(x_S-v)$-plane, points corresponding to pure crystal phases can be joined by straight lines which give the volume per particle of their coexistences. Hence, to find the set of stable phases for a given size ratio, we draw a common tangent construction as depicted in Fig. \ref{fig:common_tangent} for $q=0.23$. 
For monodisperse disks, the best packing is proven to be the hexagonal close packing \cite{toth1943}, so at $x_S=0$ and $x_S=1$ the stable phase is a hexagonal packing of large and small disks respectively.
For $x_S = 2/3$, the T1 phase achieves the best packing. The straight line that joins T1 and Hex$_L$ points gives the volume per particle of their coexistence for all compositions $x_S\in \left( 0,2/3 \right)$. This coexistence is stable because no point is found below the line. As $x_S$ is increased, S2 is found stable for $x_S = 4/5$. S2 coexists with T1 for $x_S\in \left(2/3,4/5\right)$, and with Hex$_S$ for $x_S \in \left(4/5,1\right)$.  
For clarity, the volume per particle of the coexisting hexagonal phases $v_\mathrm{Hex}(x_S) = x_S v_\mathrm{Hex_S} + (1-x_S) v_\mathrm{Hex_L}$ has been subtracted on the vertical axis of Fig. \ref{fig:common_tangent}.

Phase diagrams are mapped out by repeating this construction for the various size ratios $q$ scanned with simulations.
Once stable phases are identified from simulation snapshots, we identify contacts between pairs of particles, which provide us with a set of constraints on the particle positions, and hence define the corresponding ideal structure. The volume per particle of the ideal structures are then computed analytically as a function of the size ratio $q$ to determine their exact stability range. For this, we take into account the optimized structure of the FBMC simulations, and find an analytical solution for the particle coordinates, based on the pairs of particles in the simulated unit cell which are in direct contact after quenching. More details about this procedure can be found in the Supplementary Information. Visualisations are done using Ovito \cite{ovito}.

\section{Binary additive hard-disk mixtures}
\label{sec:HD}
\begin{figure*}
    \centering
    \includegraphics[width=0.8\linewidth]{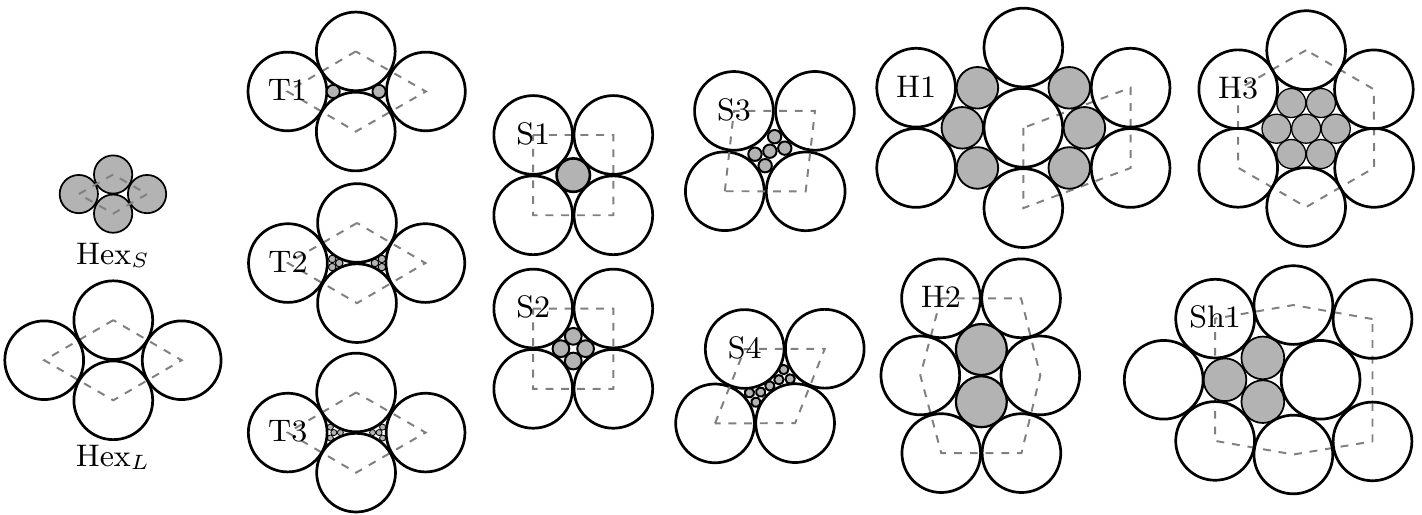}
    \caption{Stable structures that appear in the additive phase diagram. The structures a shown at their respective ''magic ratios''. Dashed grey lines outline the repeating unit of each lattice. Complete deformation paths are depicted in the SI.}
    \label{fig:stable_struct}
\end{figure*}

\begin{figure*}
    \centering
    \includegraphics[width=0.75\linewidth]{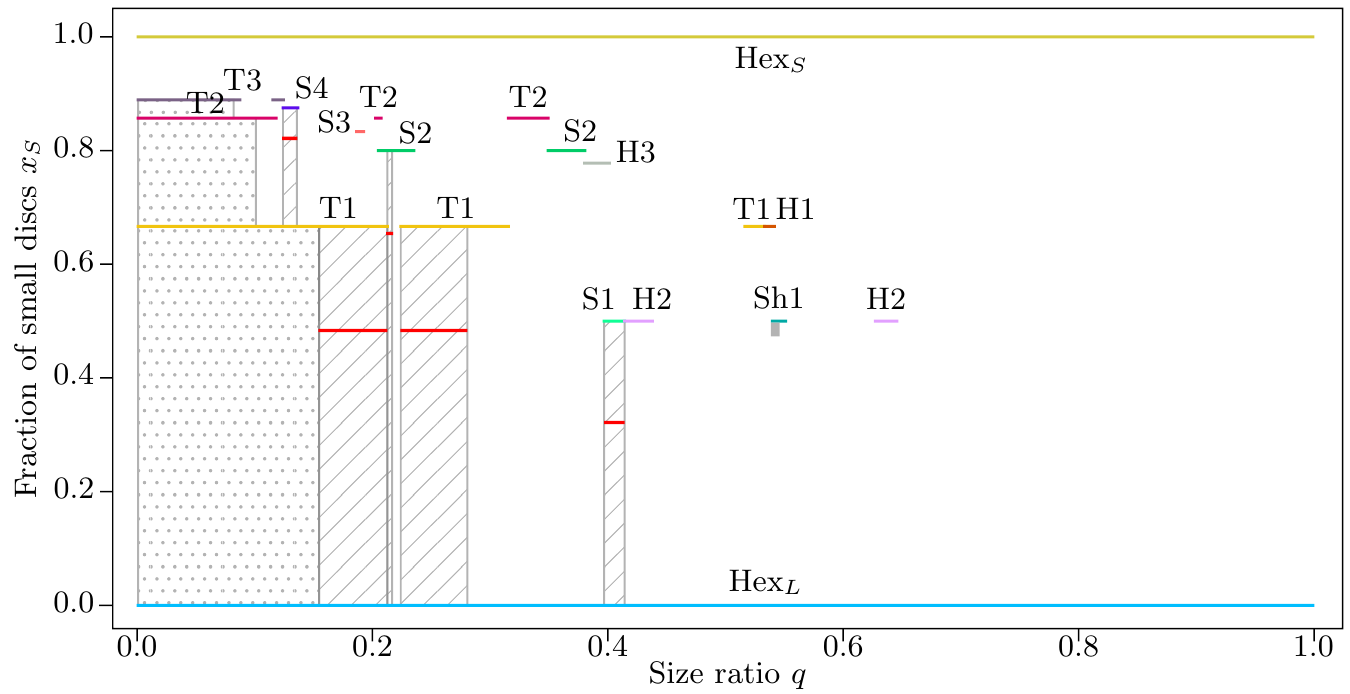}
    \caption{Infinite pressure phase diagram of binary additive hard-disk mixtures. Horizontal lines represent stability ranges of pure crystal phases. For state points outside of those lines, the stable phase is a coexistence of the two pure crystal phases that lie immediately above and below the point. The grey rectangle under Sh1 corresponds to a region where an infinite family of periodic structures are found, with the same volume per particle as the Sh1-Hex$_L$ coexistence. Dotted and hashed rectangles depict random lattice gas and random tiling regions respectively. In the latter, the horizontal red lines highlight random-tiling quasicrystals. Their compositions are $\sqrt{3}/(2+2\sqrt{3})\approx0.317$, $2\sqrt{3}/(2+3\sqrt{3})\approx0.481$, $4\sqrt{3}/(2+5\sqrt{3})\approx0.650$ and $(4+7\sqrt{3})/(6+8\sqrt{3})\approx0.812$ for S1-Hex$_L$, T1-Hex$_L$, S2-Hex$_L$ and S4-T1 random tiling quasicrystal respectively.}
    \label{fig:HD_phase_diag}
\end{figure*}

We performed the analysis outlined above for additive hard disks with size ratios between 0.05 and 1, with a step size of 0.01.
An overview of all stable crystal structures obtained -- i.e. those that correspond to the best packing for some combination of $x_S$ and $q$ -- is shown in Fig. \ref{fig:stable_struct}. Each structure is named according to the same scheme as the one used in Ref. \onlinecite{likos1993}, extended when necessary. Note that here we show each structure at a so-called ``magic'' size ratio, where a large number of neighbor pairs exactly touch. However, each binary crystal structure exists over a range of different size ratios $q$. Depending on the exact size ratio, certain``bonds'' -- or contacts between particles can be broken, and the unit cell  deformed accordingly. As an example, in the S2 phase in Fig. \ref{fig:common_tangent}, the square unit cell is slightly deformed with respect to the ``ideal'' one in Fig. \ref{fig:stable_struct}, such that some of the large particles are no longer touching. In the Supplemental Information (SI), we show for each structure how it gets deformed and which bonds are broken as the size ratio moves away from the \textit{magic} values.

To summarize the best packed structures at each size ratio and composition, we show in Fig.  \ref{fig:HD_phase_diag} the infinite-pressure phase diagram for this system.
Horizontal lines correspond to the stability ranges of pure crystal phases, which only exist at one fixed composition each. 
Points outside of those line correspond to coexistence regions of the two phases that lie directly above and below the point $(q,x_S)$.

As expected \cite{blind1969}, no stable phase other than the coexistence of two hexagonal compact packing is found for size ratios above $0.74$.
For smaller size ratios, a wealth of crystal structures are obtained.
The main features of Likos and Henley's phase diagram \cite{likos1993} are reproduced, but new stable phases (S3, S4, Sh1) are found among the candidates generated by FBMC simulations. 
S3 and S4 are both rhombic phases, with 5 and 7 small particles in the unit cell, respectively.

The repeating unit of the Sh1 lattice can be decomposed into a shield tile (hence the \emph{Sh} label), containing the 3 small disks, and 2 Hex$_L$ triangular tiles. 
As long as the size ratio is smaller than a magic ratio for which the deformed shield looses contacts between large disks, shields can be combined with Hex$_L$ triangle without volume-per-particle cost. 
Periodic structures can be constructed, that have the same volume per particle as the Sh1-Hex$_L$ coexistence \cite{fernique2019}. Examples of such structures are shown in the SI. In principle, these phases are all equally stable as the coexistence between Sh1 and Hex$_L$ at infinite pressure. At finite pressure, it is likely that vibrational entropy breaks this stalemate in favor of one specific crystal structure. However, at this point, we make no strong claims about the exact phase to be expected in this region (the very small area shaded gray in Fig.~\ref{fig:HD_phase_diag}), except that it will consist of shields and triangles. 

There are two other special regions in the phase diagram. The first, shown as dotted regions, are {\em random lattice gases}, and the second, hashed, regions are {\em random tilings}.
At small size ratios, large disks form a hexagonal compact packing ( Hex$_L$) whose interstices can host small disks. The T1, T2, and T3 phases all consist of this same hexagonal packing, in which {\it all} interstices are filled with 1,3, and 4 particles, respectively. In principle, more of these types of structures exist with more small particles in each hole\cite{uche2004}, but we have not investigated such extremely asymmetric size ratios and compositions. Where these phases coexist, it is often possible to randomly fill the interstices with a fluctuating number of particles, such that the overall composition requirement is satisfied. As this random distribution is entropically favored, such a homogeneous {\em random lattice gas} state is expected to be stable over a purely phase separated regime \cite{likos1993}. We indicate this as dotted regions in the phase diagram. Note that when the size ratio becomes too large to accommodate the small particles without deforming the hexagonal lattice, this lattice gas phase is no longer optimal in terms of packing. This can be seen, on the right edge of the lattice gas regions connecting T1 and T2, or T2 and T3. 

\begin{figure*}
    \centering
    \includegraphics[width=0.95\linewidth]{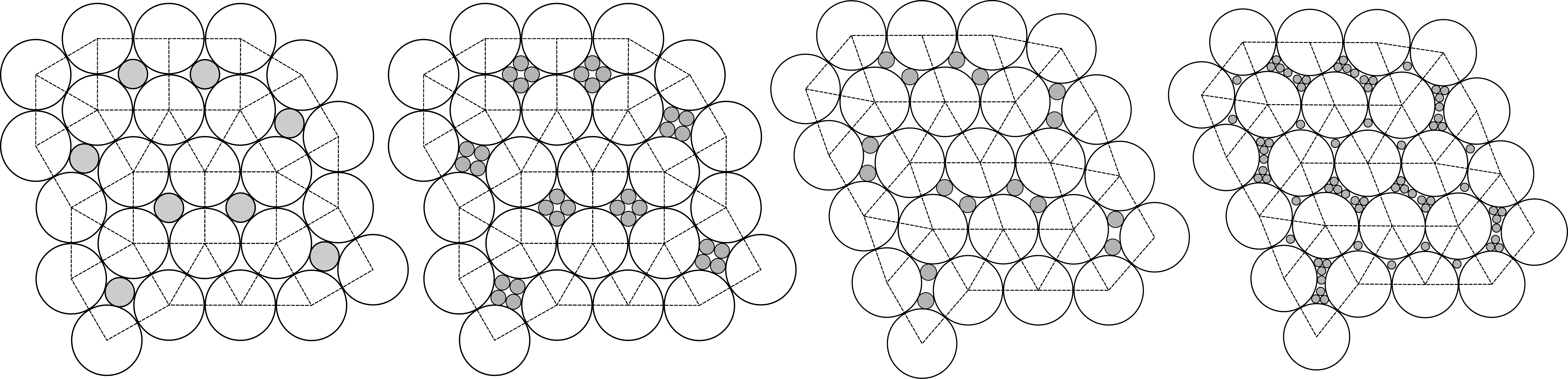}
    \caption{S1 or S2 squares and Hex$_L$ triangles can be joined without volume-per-particle cost. Moreover, squares and triangles tile the plane, so S1-Hex$_L$ and S2-Hex$_L$ coexistences result in a square-triangle random tiling (left). Random tilings can also be obtained by mixing rhombi with triangles (see for example T1-Hex$_L$ and S4-T1 random tilings (right)).}
    \label{fig:square_triangle}
\end{figure*}

The random tilings occur when the unit cells of two coexisting phases are commensurate, such that they can randomly mix. Usually, creating a boundary between two coexisting phases carries a volume cost. This normally limits mixing of phases in the infinite pressure limit, resulting in a true phase separation. However, some structures have matching unit-cell edges.
If, moreover, the structures shapes can tile the plane without gaps or overlaps, the two phases in coexistence can dissolve into one another and form a random tiling phase.
For example, as illustrated in Fig. \ref{fig:square_triangle}, there is no volume-per-particle cost for creating a boundary between S1 squares and Hex$_L$ triangles (half a Hex$_L$ unit cell), and one can tile the plane with squares and triangles. 
Random tiling and fully phase separated mixtures at the same composition $p$ pack equally well (they have the same volume per particle), however the former has a finite configurational entropy per disk \cite{widom1993}. Therefore, wherever possible, random tiling regions, depicted as hashed rectangles in the phase diagram, should be preferred, on thermodynamic grounds, compared with  phase separated coexistences. 

\begin{figure*}
    \centering
    \includegraphics[width=0.75\linewidth]{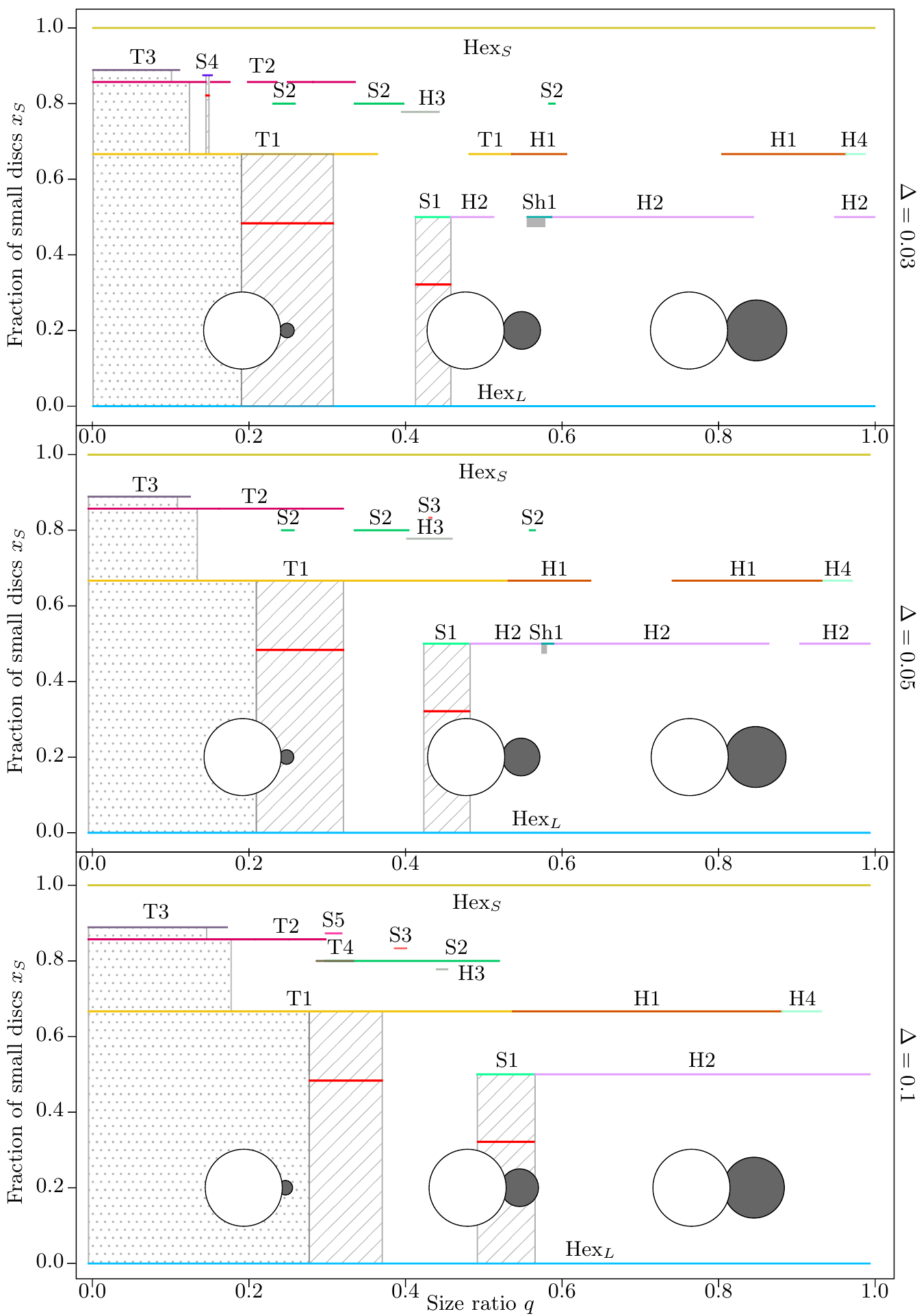}
    \caption{Infinite pressure phase diagrams of binary non-additive hard disk mixtures for $\Delta = 0.03$ (top), $\Delta = 0.05$ (middle) and $\Delta = 0.1$ (bottom). The overlap between large and small discs allowed by the non-additivity is represented in each case for $q=0.2, 0.5 \text{ and } 0.8$.}
    \label{fig:NAHD_phase_diag}
\end{figure*}

A square-triangle random tiling is an ensemble of tilings of the infinite plane with squares and triangles. As the proportion of small disks $x_S$ varies, the ratio of the number of squares $N_{sq}$ and triangles $N_{tr}$ changes. 
When $N_{sq}/N_{tr}=\sqrt{3}/4$, the random tiling ensemble has maximum entropy (the number of possible configurations is the highest) and forms a \emph{random-tiling quasicrystal} of 12-fold symmetry \cite{widom1993,kawamura1983,kalugin1994,nienhuis1998}. The corresponding compositions have been marked in red in Fig. \ref{fig:HD_phase_diag} for each random tiling region.
It has been argued that an average over this ensemble exhibits quasi-long-range order with algebraically decaying diffraction peaks at the positions of the 12-fold symmetric Bragg peaks of a quasicrystal \cite{likos1993}.

Note that the random tilings in Fig. \ref{fig:HD_phase_diag} are only square-triangle tilings when the size ratio $q$ exactly corresponds to the magic ratio for either S2 or S1. In all other cases, the squares are deformed into rhombi. The resulting random tiling is a continuous deformation of a square-triangle tiling, but no longer possesses its 12-fold symmetry. Note, for example, that random tiling pieces displayed in Fig. \ref{fig:square_triangle} are isomorphic.

The coexistence of S4 and T1 yields a new rhombus-triangle random tiling, with an associated quasicrystal. 
The FBMC simulations also revealed more optimally packed deformation paths for T2, T3 and S2 phases, that modify the extent of the stability regions in their vicinity. 
In particular, the coexistence of the new S2 deformation with Hex$_L$ is more stable than T1 at $x_S=2/3$, revealing a narrow rhombus-triangle random tiling region and hence a quasicrystal. In total, we find 4 different types of random tiling quasicrystal regions, obtained from  S1-Hex$_L$, T1-Hex$_L$, S2-Hex$_L$ and S4-T1 coexistences, at compositions $\sqrt{3}/(2+2\sqrt{3})\approx0.317$, $2\sqrt{3}/(2+3\sqrt{3})\approx0.481$, $4\sqrt{3}/(2+5\sqrt{3})\approx0.650$ and $(4+7\sqrt{3})/(6+8\sqrt{3})\approx0.812$ respectively.
Deformation paths of all the stable phases can be found in the SI.

We would like to point out that we have only investigated size ratios $q \ge 0.05$, and compositions below $x_S \le 11/12 \simeq 0.917$. This likely leads to some missed structures in the top left corner of the phase diagram. In this regime, we expect that the phase diagram gets more and more complicated for more extreme size ratios and large fractions of small disks\cite{uche2004}.  Moreover, exploring this is computationally expensive (due to large unit cells), and not necessarily likely to include interesting results. As such, we have avoided this regime of the phase diagram.

\section{Binary non-additive hard disk mixtures}
\label{sec:NAHD}
We now turn our attention to non-additive binary hard-disk mixtures, focusing on non-additivity parameters $\Delta=0.03, 0.05 \text{ and } 0.1$.
The corresponding phase diagrams are presented in Fig. \ref{fig:NAHD_phase_diag}.

\begin{figure}
    \centering
    \includegraphics[width=0.65\linewidth]{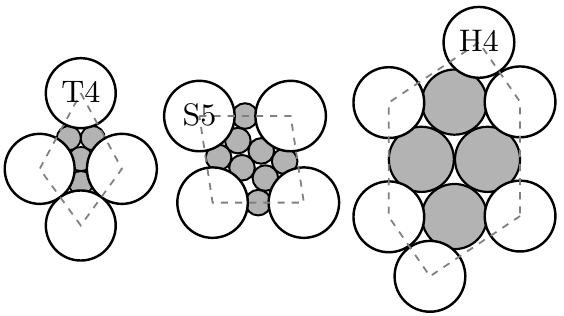}
    \caption{Repeating units of the T4, S5 and H4 lattices at $(q=0.344, \Delta=0.1)$, $(q=0.337, \Delta=0.1)$ and $(q=0.905, \Delta=0.05)$ respectively. These structures are only stable for non-additive hard disks.}
    \label{fig:NAHD_phases}
\end{figure}

One of the most immediate effects of non-additive occurs on the right-hand side of the phase diagram. While for additive disks, this region is dominated by a phase separation between large disks and small disks hexagonal crystals, non-additivity allows denser packings for high size ratios.
One of these phases, H4, was not observed at all in the additive case. The repeating unit of this lattice is presented in Fig. \ref{fig:NAHD_phases}-right. The others can be seen as variations of the H1 and H2 phases, deformed such that the lattice is approximately a hexagonal crystal of small disks with part of the particles replaced by large disks. At the exact size ratio where the contact distance between a large and a small disk ($(1-\Delta) (\sigma_S + \sigma_L) / 2$) is equal to $\sigma_S$, the large spheres can be placed randomly inside the hexagonal crystal of small spheres with no additional volume cost, leading to another zone of lattice gas. However, for values of $q$ slightly away from this magic ratio, deformations of the hexagonal lattice make this random placement unfavorable and the best-packed crystal remains periodic.

In addition to the changes at high values of $q$, two new phases, T4 and S5, are found stable at smaller size ratios. These lattices are depicted in Fig.~\ref{fig:NAHD_phases}. T4 and H4 cannot exist without non-additivity, while S5 can, but turns out to not pack efficiently enough to be stable in the additive case.

In Fig.~\ref{fig:NAHD_phases},  two global trends are observed as $\Delta$ is increased.
First, most phases can be seen as small disks enclosed into shells of large ones (see Fig. \ref{fig:stable_struct}). These phases quickly become unstable as $q$ increases beyond the point where the (cluster of) small spheres fit into the holes left by the large ones. Non-additivity mitigates the inflation of the small disk clusters as $q$ grows, which causes an overall shift of the phase diagram towards larger size ratios. 
Second, non-additivity  favors phases with a large number of contacts between large and small disks, such as T1, S1, S2, H1 and H2. Those phases gradually take over larger and larger portions of the phase diagram.

\begin{figure*}
    \centering
    \includegraphics[width=0.85\linewidth]{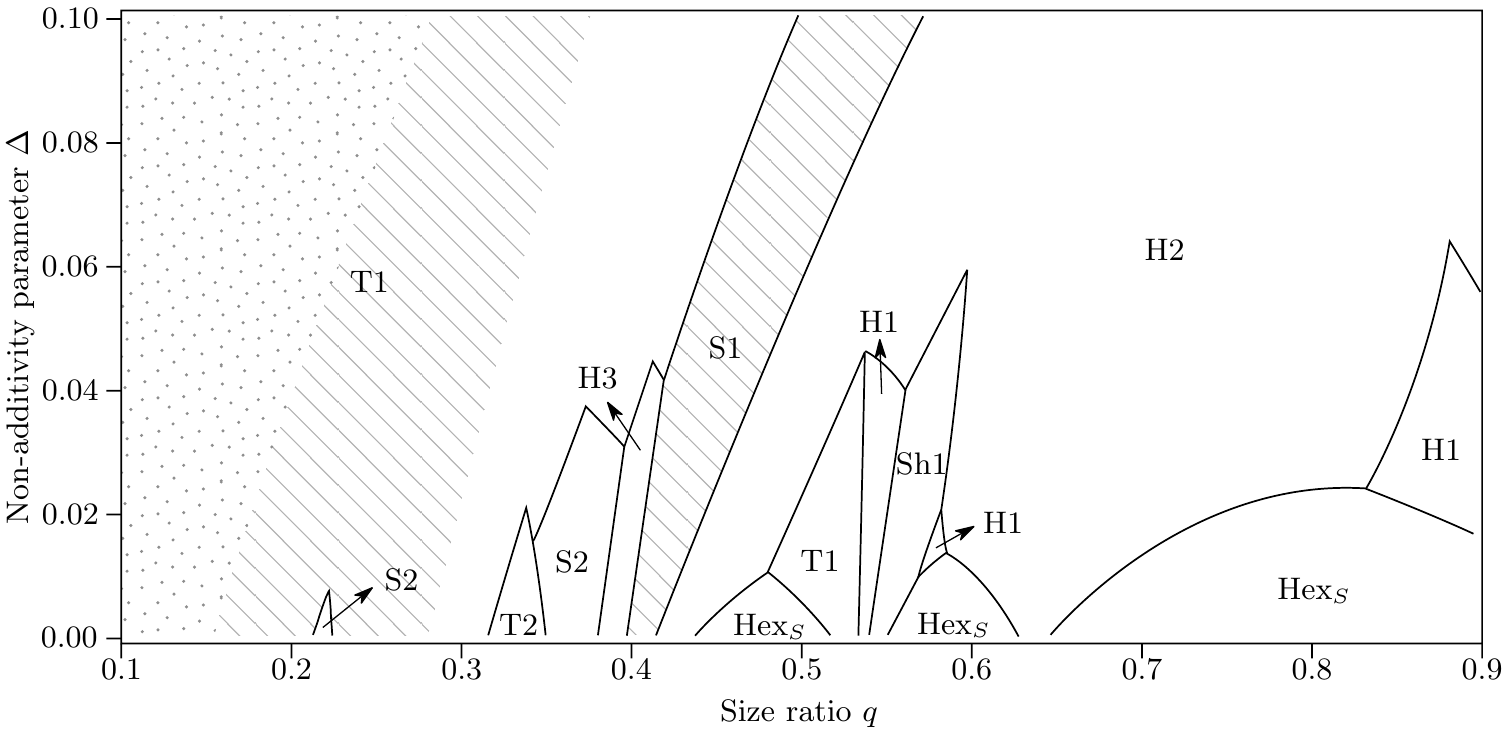}
    \caption{Evolution of the phase diagram as a function of the non-additivity parameter $\Delta \in[0,0.1]$ at fixed composition $x_S = \sqrt{3}/(2+2\sqrt{3}) \approx 0.317$ where the S1+Hex$_L$ quasicrystal is expected. At this composition, no pure periodic phase is stable, and the phase diagram consists of coexistences between Hex$_L$ and other phases. For clarity, we only display the name of the pure phase in coexistence with Hex$_L$ in the labels. As in previous diagrams, the dotted region highlight the random lattice gas and random tiling regions are hashed.}
    \label{fig:quasistability}
\end{figure*}

Another interesting effect of non-additivity is the tendency  to promote random lattice gas and random tiling regions. In Fig. \ref{fig:quasistability}, we plot the evolution of the phase diagram with $\Delta$ at a fixed composition $x_S = \sqrt{3}/(2+2\sqrt{3})\approx 0.317$ equal to the composition where quasicrystal formation is expected \cite{likos1993}.
In this way, we can, for example, follow the growth of the S1-Hex$_L$ quasicrystalline region. As $\Delta$ increases, S1 is one of the few remaining stable phases, along with T1 and H2. This results in a significant growth of the range of $q$ over which  the S1-Hex$_L$ quasicrystal is stable. In contrast, as seen in Fig. \ref{fig:NAHD_phase_diag}, the random tiling regions involving S4 and T1, and S2 and Hex$_L$ vanish for these values of $\Delta$. 

\section{Conclusions and discussion}
\label{sec:discussion}

We have systematically explored the infinite-pressure phase diagram of additive and negatively non-additive binary hard disk mixtures in two dimensions. 
These phase diagrams can serve as useful guidelines for targeted 2D self-assembly experiments, since many building blocks comprising a hard core will behave as hard particles when compressed to sufficiently high densities.
In the case of additive hard disks, our phase diagram expands on earlier work \cite{likos1993} by incorporating several new stable phases (see S3, S4, Sh1 in Fig.~\ref{fig:stable_struct}), as well as better-packed deformations of the previously identified structures. These modifications reveal two new random tiling regions (S2 in coexistence with Hex$_L$ and S4 in coexistence with T1), with their associated quasicrystals.
Hence, simple binary hard disk mixtures exhibit a surprisingly rich phase diagram, with a complexity comparable to that of three dimensional hard spheres\cite{hopkins2012}.

For the non-additive systems, we observe an overall shift of the stability regions towards larger size ratios, as well as improved stability for phases with a large number of contacts between large and small disks.
Two new phases, H4 and T4 (see Fig.~\ref{fig:stable_struct}), only possible in non-additive systems, are also found to be stable. The S5 phase, which can be constructed with additive disks but was not stable in this  case, appears as a stable phase in the non-additive mixtures. With increasing non-additivity, random tiling regions extend over larger composition ranges, potentially making them easier to observe in self-assembly experiments, where fine control over the size ratio is hard to achieve.
Negative non-additivity tends to favor contacts between large and small disks. This could also be achieved by considering selective attraction between the particles\cite{tkachenko2016}. Future studies in this direction could benefit from FBMC simulations for a systematic sampling of candidate structures.

We note that despite our systematic search for candidate crystal structures, it is impossible to exclude the possibility that additional, better packing crystal structures are possible in these systems. This is particularly relevant for the top left corner of the phase diagrams (low size ratios and high fractions of small particles), where the best-packed structures will primarily consist of structures that pack more and more small particles into the interstices between the large disks in a hexagonal lattice \cite{uche2004}. As we limit ourselves here to unit cells containing at most 12 particles, such structures are not found by our methods. We also stress that the phase diagram here is drawn at infinite pressure, where the vibrational entropy of particles can be neglected. At finite pressures, we expect the phase diagram to simplify considerably, as some structures will rapidly lose stability to other phases favored by entropic considerations. Exploration of the finite-pressure phase behavior will be the subject of a future study.

Finally, we emphasize that in the infinite pressure limit, the phase diagrams proposed here set a lower bound on the packing fraction of binary hard disks packings. Apart from 9 magic ratios for which compact packings have been demonstrated to achieve maximum packing fraction \cite{kennedy2006,fernique2020density}, it is still an open mathematical problem to prove which is the densest structure for a given composition and size ratio.

\section*{Acknowledgements}
We thank Thomas Fernique, Marianne Imp{\'e}ror-Clerc,  Jean-Fran\c{c}ois Sadoc, and Laura Filion for many useful discussions.
This work is funded by the ANR grant ANR-18-CE09-0025.

\section*{Supplementary material}
In the supplementary material, we provide representations of the deformation paths considered for the various candidate phases along with numerical values of the magic ratios for three values of the non-additivity parameter. We also discuss the existence of a family of stable periodic structures with more than 12 particles in the unit cell involving the Sh1 tile.

\section*{Data availability}
The data that support the findings of this study are available from the corresponding author upon reasonable request.

\end{document}